
  \magnification\magstep1
  \def\aa{\vskip 0.5 true cm}
  \def\bb{\vskip 0.8 true cm}
  \def\aab{\vskip 0.2 true cm}

       \font\st=cmbx12 scaled 1440
       
       \font\msym=msym10
  \hsize = 17 true cm
  \vsize = 24 true cm
  \baselineskip = 0.6 true cm
  \nopagenumbers

\overfullrule=0pt

  \def\hpm{\hbox{h$_{\pm}$}}

  \def\hmp{\hbox{h$_{\mp}$}}

  \def\upm{\hbox{u$_{\pm}$}}

  \def\ump{\hbox{u$_{\mp}$}}

  \def\ra{\hbox{$\rangle$}}

  \def\gr{\hbox{\msym R}}

  \def\de{ { {\lower 3 pt\hbox{$\scriptstyle 1$}} \over
             {\raise 3 pt\hbox{$\scriptstyle 2$}} } }
  \def\df{ { {\lower 3 pt\hbox{$\scriptstyle 3$}} \over
             {\raise 3 pt\hbox{$\scriptstyle 2$}} } }

\parindent = 1 true cm

\def\r3{{\bf r}
         \kern-0.6em{\raise 0.65em \hbox{. \kern-0.5em{.} \kern-0.5em{.}} }}
\def\z3{z\kern-0.6em{\raise 0.65em \hbox{. \kern-0.5em{.} \kern-0.5em{.}} }}
\def\y3{y\kern-0.6em{\raise 0.65em \hbox{. \kern-0.5em{.} \kern-0.5em{.}} }}
\def\x3{x\kern-0.6em{\raise 0.65em \hbox{. \kern-0.5em{.} \kern-0.5em{.}} }}

   \line{\vbox{\hsize 3.5 true cm
   \noindent
P}\hfill \vbox{\hsize 3 true cm
   \noindent
   \null\hfill \bf LYCEN 9363\break
   \null\hfill December  93}}

\vskip 2 true cm

\centerline{\st Recursion Relations for Clebsch-Gordan }

\aab
\aab
\aab

\centerline{\st Coefficients of $U_q(su_2)$ and $U_q(su_{1,1})$ }

\vskip 1.5 true cm

\centerline {M. Kibler and C. Campigotto }
\aab
\aab

\centerline{ \it Institut de Physique Nucliaire de Lyon}
\centerline{ \it IN2P3-CNRS et Universiti Claude Bernard}
\centerline{ \it 43 Bd du 11 Novembre 1918, F-69622 Villeurbanne Cedex, France}

\aa

\centerline {and}

\aa

\centerline
{Yu.F. Smirnov\footnote{$^1$}
{\it On leave of absence from the Institute of Nuclear
Physics, Moscow State University, \item{} 119899 Moscow, Russia.}}
\aab
\aab

\centerline{\it Instituto de F\'\i sica}
\centerline{\it Universidad Nacional Autonoma de M\'exico}
\centerline{\it M\'exico D.F., M\'exico}

\vskip 1.3 true cm

\centerline {\bf  Abstract}

\aa

We report in this article three- and four-term recursion relations for
Clebsch-Gordan coefficients of the quantum algebras $U_q(su_2)$ and
$U_q(su_{1,1})$. These relations were obtained by exploiting the
complementarity of three quantum
algebras in a $q$-deformation of $sp(8, \gr)$.

This article constitutes a contribution to the proceedings
of
the International workshop ``Symmetry Methods in Physics''
organized by the {\it Joint Institute for Nuclear Research}
(Dubna, Russia, 6-11 July 1993) in memory of Professor
Ya.A. Smorodinski{\u \i}.

\vfill\eject

\centerline{\st Recursion Relations for Clebsch-Gordan }

\aab
\aab
\aab

\centerline{\st Coefficients of $U_q(su_2)$ and $U_q(su_{1,1})$ }

\vskip 1.5 true cm

\centerline {M. Kibler and C. Campigotto }
\aab
\aab

\centerline{ \it Institut de Physique Nucliaire de Lyon}
\centerline{ \it IN2P3-CNRS et Universiti Claude Bernard}
\centerline{ \it 43 Bd du 11 Novembre 1918, F-69622 Villeurbanne Cedex, France}

\aa

\centerline {and}

\aa

\centerline
{Yu.F. Smirnov\footnote{$^1$}
{\it On leave of absence from the Institute of Nuclear
Physics, Moscow State University, \item{} 119899 Moscow, Russia.}}
\aab
\aab

\centerline{\it Instituto de F\'\i sica}
\centerline{\it Universidad Nacional Autonoma de M\'exico}
\centerline{\it M\'exico D.F., M\'exico}

\vskip 1.3 true cm

\centerline {\bf  Abstract}

\aa

We report in this article three- and four-term recursion relations for
Clebsch-Gordan coefficients of the quantum algebras $U_q(su_2)$ and
$U_q(su_{1,1})$. These relations were obtained by exploiting the
complementarity of three quantum
algebras in a $q$-deformation of $sp(8, \gr)$.

\vskip 1.5 true cm

\centerline {\bf 1. Introduction}

\aa

The theory  of quantum algebras has been the object of numerous investigations
both in physics and mathematics. In particular, the one-parameter quantum
algebras $U_q(su_{2})$ and $U_q(su_{1,1})$ have been investigated by many
authors
(see, for instance, Refs.~[1,2]). In addition,
two-parameter deformations of
$su_2$ and $u_2$ have been worked out in various papers [3-8].

\aab

The application to physics of the quantum algebras $U_q(su_{2})$ and
$U_q(su_{1,1})$ requires the knowledge of the corresponding coupling and
recoupling coefficients. Clebsch-Gordan coefficients for $U_q(su_{2})$ (and
$U_q(su_{1,1})$), in one- and two-parameter formulations, have been calculated
by several people (e.g., see Refs.~[5,7]). In addition, recursion relations for
Clebsch-Gordan coefficients of $U_q(su_{2})$ and $U_q(su_{1,1})$ have also been
derived,
 as an extension of the relations corresponding to the
$q=1$ case [9-13],
in Refs.~[14-16].

\aab

It is the aim of this contribution to list three- and four-term
recursion
relations for the Clebsch-Gordan coefficients of $U_q(su_{2})$ derived from
an algorithm recently
described in Ref.~[17]. We shall also give a few three-term
recursion
relations for the Clebsch-Gordan coefficients of $U_q(su_{1,1})$ derived by
means of this algorithm.

\aab

This work is dedicated to the memory of the late Professor
Ya.A. Smorodinski{\u \i} who contributed, among many other
fields, to the Wigner-Racah algebra of the group $SU_2$.

\bb

\centerline {\bf 2. The algorithm}

\aa

The algorithm applied in this article
follows from the work of Schwinger [9]
and was implicitly used by Rasmussen [11]
and Kibler and Grenet [12] in order to
derive recursion relations in the $q=1$ case.
In the $q \ne 1$ case, this algorithm
has been described in detail by Smirnov and
Kibler [17]. We shall not repeat
the description of the algorithm
in the present paper. However, we shall
briefly mention here its main features.

\aab

We start from a $q$-deformation of the dynamical invariance algebra $sp(8,\gr)$
of the four-dimensional harmonic oscillator. Following the work of Moshinsky
and Quesne [18] on the complementarity of Lie groups, we can extract three
complementary algebras from two chains of quantum algebras having the
$q$-deformation of $sp(8,\gr)$ as the head algebra. These algebras are denoted
in Ref.~[17] as $U_q (su_{2}^{\cal J})$,  $U_q (su_{2}^{\Lambda})$ and
 $U_q (su_{1,1}^{\cal K})$ and their generators are collectively indicated by
${\cal J }$,
${\Lambda}$ and
${\cal K }$, respectively. The
latter generators are built from
the four pairs of $q$-boson
operators and the four number
operators corresponding to the four-dimensional $q$-deformed harmonic
oscillator. They are defined in such a way to satisfy the co-product rules for
the Hopf algebras  $U_q (su_{2}^{\cal J})$,  $U_q (su_{2}^{\Lambda})$ and
 $U_q (su_{1,1}^{\cal K})$ (see Ref.~[17]).

\aab

The algorithm amounts to calculating, in two different ways,
matrix elements of the type
$\langle n_1 n_2 n_3 n_4 | X | j : \mu m \kappa \rangle$, where $X$ is either
a linear form or a bilinear form of the generators
$\{ {\cal  J} \}$,
$\{ {\Lambda} \}$ and
$\{ {\cal  K} \}$. Furthermore, the vectors
$| n_1 n_2 n_3 n_4 \rangle$ are state vectors
for the four-dimensional harmonic
oscillator and the vectors
$| j : \mu m \kappa \rangle$ are common eigenstates
of ${\cal J}^2$, ${\cal J}_3$, $\Lambda_3$ and ${\cal K}_3$, where ${\cal J}^2$
stands for the common Casimir operator of $U_q (su_{2}^{\cal J})$,
$U_q (su_{2}^{\Lambda})$ and  $U_q (su_{1,1}^{\cal K})$.

\aab

It should be emphasized that the algorithm just described also furnishes an
elegant way for deriving the $q$-analogue of Regge symmetries for $U_q
(su_{2})$
as well as some connecting  formulas between Clebsch-Gordan coefficients of
$U_q (su_{2})$ and $U_q (su_{1,1})$ [17].

\bb

\centerline {\bf 3. Recursion relations}

\aa

We give below recursion relations for the Clebsch-Gordan coefficients of
$U_q (su_{2})$ and  $U_q (su_{1,1})$ obtained according to the algorithm
described in Section 2.

\aab

In what follows, we employ the abbreviations
$$
\eqalign
{
          [x] \ & = \ {q^x - q^{-x} \over q - q^{-1} } \cr
  {\rm h}_{\pm} \ = \ \de & \pm \de \quad
  {\rm h}_{\mp} \ = \ \de   \mp \de                    \cr
  {\rm u}_{\pm} \ = \ 1   & \pm 1   \quad
  {\rm u}_{\mp} \ = \ 1     \mp 1                      \cr
}
$$
In addition, the Clebsch-Gordan coefficients for $U_q (su_{1,1})$ (for the
positive discrete series) and $U_q (su_{2})$ are written as
$(k_1 k_2 \kappa_1 \kappa_2 | j \kappa)_q$  and
$(j_1 j_2 m_1      m_2      | j m     )_q$, respectively.

\aa

\noindent {\it 3.1. Recursion relations for $U_q(su_2)$}

\aa

1. The action of ${\cal J}_{\pm}$ on
$\vert j_1 \mp \de,
       j_2 \pm \de, j, m \ra_q $
leads to

$$
\eqalign
{
&\sqrt{ [j \mp j_1 \pm j_2 + 1]   \
        [j \pm j_1 \mp j_2    ] } \
                           (j_1 j_2 m_1 m_2 \vert j m )_{q} \cr
= & q^{ + {1 \over 2} (j_1 - m_1 - j_2 + m_2) } \
\sqrt{ [j_1 + m_1 + \hmp]   \
       [j_2 + m_2 + \hpm] } \
 (j_1 \mp \de, j_2 \pm \de, m_1 \mp \de, m_2 \pm \de \vert j m )_{q} \cr
+ & q^{ - {1 \over 2} (j_1 + m_1 - j_2 - m_2) } \
\sqrt{ [j_1 - m_1 + \hmp ]   \
       [j_2 - m_2 + \hpm ] } \
 (j_1 \mp \de, j_2 \pm \de, m_1 \pm \de, m_2 \mp \de \vert j m )_{q} \cr
}
$$

2. The action of ${\cal K}_{\pm}$ on
$\vert j_1 \mp \de,
       j_2 \mp \de, j, m \ra_q $
leads to

$$
\eqalign
{
& \sqrt{ [j_1 + j_2 - j +     \hmp]   \
         [j_1 + j_2 + j + 1 + \hmp] } \
                           (j_1 j_2 m_1 m_2 \vert j m )_{q} \cr
= & q^{ + {1 \over 2} (j_1 - m_1 + j_2 + m_2 + 1) } \
\sqrt{ [j_1 + m_1 + \hmp]   \
       [j_2 - m_2 + \hmp] } \
(j_1 \mp \de, j_2 \mp \de, m_1 \mp \de, m_2 \pm \de \vert j m )_{q} \cr
- & q^{ - {1 \over 2} (j_1 + m_1 + j_2 - m_2 + 1)} \
\sqrt{ [j_1 - m_1 + \hmp]   \
       [j_2 + m_2 + \hmp] } \
(j_1 \mp \de, j_2 \mp \de, m_1 \pm \de, m_2 \mp \de \vert j m )_{q} \cr
}
$$

3. The action of ${\Lambda}_{\pm}$ on
$\vert j_1, j_2, j, m \mp 1 \ra_q $
leads to

$$
\eqalign
{
 \sqrt{ [j \pm m] \ [j \mp m + 1] } \ & ( j_1 j_2 m_1 m_2 \vert j m )_{q} \cr
 = & \ q^{+ m_2} \ \sqrt{ [j_1 \mp m_1 + 1] \ [j_1 \pm m_1]} \
        (j_1, j_2, m_1 \mp 1, m_2 \vert j m \mp 1 )_{q} \cr
 + & \ q^{- m_1} \ \sqrt{ [j_2 \mp m_2 + 1] \ [j_2 \pm m_2]} \
        (j_1, j_2, m_1, m_2 \mp 1 \vert j m \mp 1 )_{q} \cr
}
$$

\aa

In the limiting case where $q=1$, the recursion relations 1 to
3 give back well-known relations for the Lie algebra $su_2$
(see Refs.~[12,13]).
The six preceding recursion relations are in accordance with
the results by Nomura [14],
   Groza {\it et al.} [15],
  Kachurik and Klimyk [15], and
               Aizawa [16]
who derived recursion relations for the Clebsch-Gordan
coefficients of $U_q(su_2)$
from $q$-deformed hypergeometric functions.

\aab

We now give twenty four-term recursion relations
that can be obtained from twenty operators of
type
${\cal K}  {\cal K }$,
${\cal J}  {\cal J }$,
${\cal J}  {\cal K }$,
${\Lambda} {\Lambda}$,
${\Lambda} {\cal K }$ and
${\Lambda} {\cal J }$.

\aa

4. The action of ${\cal K}_{\pm}
                  {\cal K}_{\pm}$ on
       $\vert j_1 \mp 1,
              j_2 \mp 1, j, m \ra_q $
leads to

$$
\eqalign
{
& \sqrt{ [j_1 + j_2 + j + 1 + \ump] \
         [j_1 + j_2 + j + \ump    ] \
         [j_1 + j_2 - j + \ump    ] \
         [j_1 + j_2 - j \mp 1     ] } \cr
& (j_1 j_2 m_1 m_2 \vert j m )_{q} \cr
= & \ q^{+ j_1 - m_1 + j_2 + m_2 + 1}
 \sqrt{ [j_1 + m_1 \mp 1 ] \
        [j_1 + m_1 + \ump] \
        [j_2 - m_2 \mp 1 ] \
        [j_2 - m_2 + \ump] }
\cr
& {\phantom  { \ q^{+ j_1 - m_1 + j_2 + m_2 + 1}} }
(j_1 \mp 1,
 j_2 \mp 1,
 m_1 \mp 1,
 m_2 \pm 1 \vert j m )_{q} \cr
- & \ [2] \ q^{m_2 - m_1}
 \sqrt{ [j_1 + m_1 + \hmp] \
        [j_1 - m_1 + \hmp] \
        [j_2 + m_2 + \hmp] \
        [j_2 - m_2 + \hmp] }
\cr
& {\phantom { \ q^{+ j_1 - m_1 + j_2 + m_2 + 1}} }
                (j_1 \mp 1,
                 j_2 \mp 1, m_1, m_2 \vert j m )_{q} \cr
+ & \ q^{- j_1 - m_1 - j_2 + m_2 - 1}
 \sqrt{ [j_1 - m_1 \mp 1 ] \
        [j_1 - m_1 + \ump] \
        [j_2 + m_2 \mp 1 ] \
        [j_2 + m_2 + \ump] }
\cr
&  {\phantom { \ q^{+ j_1 - m_1 + j_2 + m_2 + 1} }}
 (j_1 \mp 1,
  j_2 \mp 1,
  m_1 \pm 1,
  m_2 \mp 1 \vert j m )_{q} \cr
}
$$

5. The action of ${\cal J}_{\pm}
                  {\cal J}_{\pm}$ on $\vert j_1 \mp 1,
                                            j_2 \pm 1, j, m \ra_q $
leads to

$$
\eqalign
{
& \sqrt{ [j \mp j_1 \pm j_2 + 2] \
         [j \pm j_1 \mp j_2 - 1] \
         [j \mp j_1 \pm j_2 + 1] \
         [j \pm j_1 \mp j_2    ] } (j_1 j_2 m_1 m_2 \vert j m )_{q} \cr
= &            \ q^{+ j_1 - m_1 - j_2 + m_2}
 \sqrt{ [j_1 + m_1 \mp 1 ] \
        [j_1 + m_1 + \ump] \
        [j_2 + m_2 \pm 1 ] \
        [j_2 + m_2 + \upm] } \cr
&  {\phantom { \ q^{+ j_1 - m_1 - j_2 + m_2}} }
(j_1 \mp 1,
 j_2 \pm 1,
 m_1 \mp 1,
 m_2 \pm 1 \vert j m )_{q} \cr
+ & \ [2] \ q^{m_2 - m_1}
 \sqrt{ [j_1 - m_1 + \hmp] \
        [j_1 + m_1 + \hmp] \
        [j_2 - m_2 + \hpm] \
        [j_2 + m_2 + \hpm] } \cr
&  {\phantom { \ q^{+ j_1 - m_1 - j_2 + m_2}} }
(j_1 \mp 1,
 j_2 \pm 1, m_1, m_2 \vert j m )_{q} \cr
+ & \ q^{- j_1 - m_1 + j_2 + m_2}
 \sqrt{ [j_1 - m_1 \mp 1 ] \
        [j_1 - m_1 + \ump] \
        [j_2 - m_2 \pm 1 ] \
        [j_2 - m_2 + \upm] } \cr
& {\phantom { \ q^{+ j_1 - m_1 - j_2 + m_2}} }
(j_1 \mp 1,
 j_2 \pm 1,
 m_1 \pm 1,
 m_2 \mp 1 \vert j m )_{q} \cr
}
$$

6. The action of ${\cal J}_{\pm}
                  {\cal K}_{\pm}$ on
$\vert j_1 \mp 1, j_2, j, m \ra_q $
leads to

$$
\eqalign
{
& \sqrt{ [j_1 +   j_2 \mp j + \ump] \
         [j_1 +   j_2 \pm j + 1   ] \
         [j   \pm j_1 \mp j_2     ] \
         [j   \mp j_1 \pm j_2 + 1 ] }
( j_1 j_2 m_1 m_2 \vert j m )_{q} \cr
& = \ q^{+ j_1 - m_1 + m_2 + \hpm}
  \sqrt{ [j_1 + m_1 + \ump]  \
         [j_1 + m_1 \mp 1 ]  \
         [j_2 \mp m_2     ]  \
         [j_2 \pm m_2 + 1 ]} \cr
& {\phantom {= \ q^{+ j_1 - m_1 + m_2 + \hpm} }}
  ( j_1 \mp 1, j_2,
    m_1 \mp 1,
    m_2 \pm 1 \vert j m )_{q} \cr
& + \ q^{m_2 - m_1}
 \sqrt{ [j_1 + m_1 + \hmp]  \
        [j_1 - m_1 + \hmp]} \cr
&
\left\{   [j_2 - m_2 + \hpm] \ q^{+ j_2 + \hmp}
        - [j_2 + m_2 + \hpm] \ q^{- j_2 - \hmp} \right\}
( j_1 \mp 1, j_2, m_1, m_2 \vert j m )_{q} \cr
& - \ q^{- j_1 - m_1 + m_2 - \hpm}
 \sqrt{ [j_1 - m_1 + \ump] \
        [j_1 - m_1 \mp 1 ] \
        [j_2 \pm m_2     ] \
        [j_2 \mp m_2 + 1 ]} \cr
& {\phantom {= \ q^{+ j_1 - m_1 + m_2 + \hpm} }}
  ( j_1 \mp 1, j_2,
    m_1 \pm 1,
    m_2 \mp 1 \vert j m )_{q} \cr
}
$$

7. The action of ${\cal J}_{\pm}
                  {\cal K}_{\mp}$ on
$\vert j_1, j_2 \pm 1, j, m \ra_q $
leads to

$$
\eqalign
{
& \sqrt{ [j_1 + j_2 \pm j   + \upm] \
         [j_1 + j_2 \mp j   + 1   ] \
         [j \mp j_1 \pm j_2 + 1   ] \
         [j \pm j_1 \mp j_2       ] }
(j_1 j_2 m_1 m_2 \vert j m )_{q} \cr
= & - q^{-j_2 + m_2 - m_1 - \hmp}
 \sqrt{ [j_1 \pm m_1     ] \
        [j_1 \mp m_1 + 1 ] \
        [j_2 + m_2 + \upm] \
        [j_2 + m_2 \pm 1]} \cr
&  {\phantom { \ q^{-j_2 + m_2 - m_1}} }
(j_1, j_2 \pm 1, m_1 \mp 1, m_2 \pm 1 \vert j m )_{q} \cr
& + q^{m_2 - m_1}
 \sqrt{ [j_2 + m_2 + \hpm]  \
        [j_2 - m_2 + \hpm]} \cr
& \left\{ [j_1 + m_1 + \hmp] \ q^{+j_1 + \hpm} -
          [j_1 - m_1 + \hmp] \ q^{-j_1 - \hpm} \right\} \cr
&  {\phantom { \ q^{-j_2 + m_2 - m_1}} }
(j_1, j_2 \pm 1, m_1, m_2 \vert j m )_{q} \cr
& + q^{+j_2 + m_2 - m_1 + \hmp}
 \sqrt{ [j_1 \mp m_1     ] \
        [j_1 \pm m_1 + 1 ] \
        [j_2 - m_2 + \upm] \
        [j_2 - m_2 \pm 1 ]} \cr
&  {\phantom { \ q^{-j_2 + m_2 - m_1}} }
(j_1, j_2 \pm 1, m_1 \pm 1, m_2 \mp 1 \vert j m )_{q} \cr
}
$$

8. The action of ${\Lambda}_{\pm}
                  {\Lambda}_{\pm}$ on
$\vert j_1, j_2, j, m \mp 2 \ra_q $
leads to

$$
\eqalign
{
& \sqrt{ [j \pm m - 1] \
         [j \pm m    ] \
         [j \mp m + 1] \
         [j \mp m + 2] }
( j_1 j_2 m_1 m_2 \vert j m )_{q} \cr
= & \ q^{+ 2 m_2}
 \sqrt{ [j_1 \pm m_1 - 1]  \
        [j_1 \pm m_1    ]  \
        [j_1 \mp m_1 + 1]  \
        [j_1 \mp m_1 + 2]} \cr
&  {\phantom { \ q^{+2 m_2}} }
( j_1, j_2, m_1 \mp 2, m_2 \vert j,
            m   \mp 2 )_{q} \cr
+ & \ [2] \ q^{m_2 - m_1}
 \sqrt{ [j_1 \pm m_1    ]  \
        [j_1 \mp m_1 + 1]  \
        [j_2 \pm m_2    ]  \
        [j_2 \mp m_2 + 1]} \cr
&  {\phantom { \ q^{+2 m_2}} }
( j_1, j_2, m_1 \mp 1,
            m_2 \mp 1 \vert j,
            m   \mp 2 )_{q} \cr
+ & \ q^{- 2 m_1}
 \sqrt{ [j_2 \pm m_2 - 1]  \
        [j_2 \pm m_2    ]  \
        [j_2 \mp m_2 + 1]  \
        [j_2 \mp m_2 + 2]} \cr
&  {\phantom { \ q^{+2 m_2}} }
( j_1, j_2, m_1, m_2 \mp 2 \vert j,
                 m   \mp 2 )_{q} \cr
}
$$

9. The action of ${\Lambda}_+
                  {\cal J}_{\pm}$ on
$\vert j_1 \mp \de,
       j_2 \pm \de, j, m - 1 \ra_q $
leads to

$$
\eqalign
{
& \sqrt{ [j - m + 1            ] \
         [j + m                ] \
         [j \mp j_1 \pm j_2 + 1] \
         [j \pm j_1 \mp j_2    ] }
(j_1 j_2 m_1 m_2 \vert j m )_{q} \cr
= & q^{\pm{1 \over 2}(j_1 \mp m_1 - j_2 \pm 3 m_2 \pm 1)}
 \sqrt{ [j_1 \pm m_1 \mp 1   ] \
        [j_1 - m_1 + 1 + \hmp] \
        [j_1 + m_1           ] \
        [j_2 \pm m_2 + \hpm  ]} \cr
&  {\phantom { q^{\pm{1 \over 2}(j_1 \mp m_1 - j_2 \pm 3 m_2 \pm 1)}} }
(j_1 \mp \de, j_2 \pm \de, m_1 - \df, m_2 + \de \vert j, m-1 )_{q} \cr
+ & \left\{ [j_1 \mp m_1 + \hpm] \
q^{\mp{1 \over 2}(j_1 \pm m_1 - j_2 \mp 3 m_2 \mp 1)}
+           [j_2 \pm m_2 + \hmp] \
q^{\pm{1 \over 2}(j_1 \mp 3 m_1 - j_2 \pm m_2 \mp 1)} \right\} \cr
&  \sqrt{ [j_1 \pm m_1 + \hmp] \
          [j_2 \mp m_2 + \hpm] }
(j_1 \mp \de, j_2 \pm \de, m_1 - \de, m_2 - \de \vert j, m-1 )_{q} \cr
+ & q^{\mp{1 \over 2}(j_1 \pm 3 m_1 - j_2 \mp m_2 \pm 1)}
 \sqrt{
[j_2 \mp m_2 \pm 1]    \
[j_2 - m_2 + 1 + \hpm] \
[j_2 + m_2]            \
[j_1 \mp m_1 + \hmp]   } \cr
&  {\phantom { q^{\pm{1 \over 2}(j_1 \mp m_1 - j_2 \pm 3 m_2 \pm 1)}} }
(j_1 \mp \de, j_2 \pm \de, m_1 + \de, m_2 - \df \vert j, m-1 )_{q} \cr
}
$$

10. The action of ${\Lambda}_+
                   {\cal  K}_{\pm}$ on
$\vert j_1 \mp \de,
       j_2 \mp \de, j, m - 1 \ra_q $
leads to

$$
\eqalign
{
& \sqrt{ [j - m + 1             ] \
         [j + m                 ] \
         [j_1 + j_2 \pm j + 1   ] \
         [j_1 + j_2 \mp j + \ump] }
(j_1 j_2 m_1 m_2 \vert j m )_{q} \cr
=
& \pm q^{\pm{1 \over 2}(j_1 \mp m_1 + j_2 \pm 3 m_2 + \upm)}
 \sqrt{ [j_1 + m_1           ] \
        [j_1 \pm m_1 \mp 1   ] \
        [j_1 - m_1 + 1 + \hmp] \
        [j_2 \mp m_2 + \hmp  ]}\cr
&  {\phantom { \pm q^{\pm{1 \over 2}(j_1 \mp m_1 + j_2 \pm 3 m_2 + \upm)}} }
(j_1 \mp \de,
 j_2 \mp \de,
 m_1 -   \df,
 m_2 +   \de \vert j, m-1 )_{q} \cr
& \pm \left\{
  [j_2 \mp m_2 + \hpm] \ q^{\pm{1 \over 2}(j_1 \mp 3 m_1 + j_2 \pm m_2 + \ump)}
- [j_1 \mp m_1 + \hpm] \ q^{\mp{1 \over 2}(j_1 \pm m_1 + j_2 \mp 3 m_2 + \ump)}
\right\} \cr
&   \sqrt{ [j_1 \pm m_1 + \hmp] \
           [j_2 \pm m_2 + \hmp]}
(j_1 \mp \de,
 j_2 \mp \de,
 m_1  -  \de,
 m_2  -  \de \vert j, m-1 )_{q} \cr
& \mp q^{\mp{1 \over 2}(j_1 \pm 3 m_1 + j_2 \mp m_2 + \upm)}
 \sqrt{ [j_2 + m_2           ] \
        [j_2 \pm m_2 \mp 1   ] \
        [j_2 - m_2 + 1 + \hmp] \
        [j_1 \mp m_1 + \hmp  ]} \cr
&  {\phantom { \pm q^{\pm{1 \over 2}(j_1 \mp m_1 + j_2 \pm 3 m_2 + \upm)}} }
(j_1 \mp \de,
 j_2 \mp \de,
 m_1   + \de,
 m_2  -  \df \vert j, m-1 )_{q} \cr
}
$$

11. The action of ${\Lambda}_-
                   {\cal  J}_{\pm}$ on
$\vert j_1 \mp \de,
       j_2 \pm \de, j, m + 1 \ra_q $
leads to

$$
\eqalign
{
& \sqrt{ [j - m                ] \
         [j + m + 1            ] \
         [j \mp j_1 \pm j_2 + 1] \
         [j \pm j_1 \mp j_2    ] }
(j_1 j_2 m_1 m_2 \vert j m )_{q} \cr
& = q^{\pm{1 \over 2}(j_1 \mp 3 m_1 - j_2 \pm m_2 \pm 1)}
 \sqrt{ [j_2 - m_2           ]  \
        [j_2 \pm m_2 \pm 1   ]  \
        [j_2 + m_2 + 1 + \hpm]  \
        [j_1 \pm m_1 + \hmp  ] }\cr
&  {\phantom  { q^{\pm{1 \over 2}(j_1 \mp 3 m_1 - j_2 \pm m_2 \pm 1)}} }
(j_1 \mp \de,
 j_2 \pm \de,
 m_1   - \de,
 m_2  +  \df \vert j, m+1 )_{q} \cr
& + \left\{
  [j_1 \pm m_1 + \hpm] \ q^{\pm{1 \over 2}(j_1 \mp m_1 - j_2 \pm 3 m_2 \mp 1)}
+ [j_2 \mp m_2 + \hmp] \ q^{\mp{1 \over 2}(j_1 \pm 3 m_1 - j_2 \mp m_2 \mp 1)}
    \right\} \cr
&  \sqrt{ [j_1 \mp m_1 + \hmp] \
          [j_2 \pm m_2 + \hpm] }
(j_1 \mp \de,
 j_2 \pm \de,
 m_1  +  \de,
 m_2  +  \de \vert j, m+1 )_{q} \cr
& + q^{\mp{1 \over 2}(j_1 \pm m_1 - j_2 \mp 3 m_2 \pm 1)}
 \sqrt{ [j_1 - m_1           ]  \
        [j_1 \mp m_1 \mp 1   ]  \
        [j_1 + m_1 + 1 + \hmp]  \
        [j_2 \mp m_2 + \hpm  ]} \cr
&  {\phantom  { q^{\pm{1 \over 2}(j_1 \mp 3 m_1 - j_2 \pm m_2 \pm 1)}} }
(j_1 \mp \de,
 j_2 \pm \de,
 m_1  +  \df,
 m_2  -  \de \vert j, m+1 )_{q} \cr
}
$$

12. The action of ${\Lambda}_-
                   {\cal  K}_{\pm}$ on
$\vert j_1 \mp \de,
       j_2 \mp \de, j, m + 1 \ra_q $
leads to

$$
\eqalign
{
& \sqrt{ [j - m                 ] \
         [j + m + 1             ] \
         [j_1 + j_2 \mp j + \ump] \
         [j_1 + j_2 \pm j + 1   ] }
(j_1 j_2 m_1 m_2 \vert j m )_{q} \cr
=
& \pm q^{ \pm{1 \over 2}(j_1 \mp 3 m_1 + j_2 \pm m_2 + \upm)}
 \sqrt{ [j_1 \pm m_1 + \hmp] \
        [j_2 - m_2           ]   \
        [j_2 \mp m_2 \mp 1   ]   \
        [j_2 + m_2 + 1 + \hmp] } \cr
& {\phantom {  \pm q^{ \pm{1 \over 2}(j_1 \mp 3 m_1 + j_2 \pm m_2 + \upm)}} }
(j_1 \mp \de,
 j_2 \mp \de,
 m_1  -  \de,
 m_2  +  \df \vert j, m+1 )_{q} \cr
& \pm \left\{
[j_1 \pm m_1 + \hpm] \ q^{\pm{1 \over 2}(j_1 \mp m_1 + j_2 \pm 3 m_2 + \ump)}
-
[j_2 \pm m_2 + \hpm] \ q^{\mp{1 \over 2}(j_1 \pm 3 m_1 + j_2 \mp m_2 + \ump)}
 \right\} \cr
&  \sqrt{ [j_1 \mp m_1 + \hmp] \
          [j_2 \mp m_2 + \hmp] }
(j_1 \mp \de,
 j_2 \mp \de,
 m_1   + \de,
 m_2   + \de \vert j, m+1 )_{q} \cr
& \mp q^{\mp{1 \over 2}(j_1 \pm m_1 + j_2 \mp 3 m_2 + \upm)}
 \sqrt{ [j_2 \pm m_2 + \hmp  ]   \
        [j_1 - m_1           ]   \
        [j_1 \mp m_1 \mp 1   ]   \
        [j_1 + m_1 + 1 + \hmp] } \cr
& {\phantom {  \pm q^{ \pm{1 \over 2}(j_1 \mp 3 m_1 + j_2 \pm m_2 + \upm)}} }
(j_1 \mp \de,
 j_2 \mp \de,
 m_1   + \df,
 m_2   - \de \vert j, m+1 )_{q} \cr
}
$$

13. The action of ${\Lambda}_{\pm}
                   {\Lambda}_{\mp}$ on $\vert j_1, j_2, j, m \ra_q $
leads to

$$
\eqalign
{
&  [j \pm m] \ [j \mp m + 1]
(j_1 j_2 m_1 m_2 \vert j m )_{q} \cr
= & \ q^{m_2 - m_1 + 1}
 \sqrt{ [j_1 - m_1 + 1] \ [j_1 + m_1] \
        [j_2 + m_2 + 1] \ [j_2 - m_2] \ }
(j_1, j_2, m_1-1, m_2+1 \vert j m )_{q} \cr
+ &  \left\{
     q^{+2 m_2} \ [j_1 \mp m_1 + 1] \ [j_1 \pm m_1]
+    q^{-2 m_1} \ [j_2 \mp m_2 + 1] \ [j_2 \pm m_2]
     \right\}
(j_1 j_2 m_1 m_2 \vert j m )_{q} \cr
+ & \ q^{m_2 - m_1 - 1}
 \sqrt{ [j_1 + m_1 + 1] \ [j_1 - m_1] \
        [j_2 - m_2 + 1] \ [j_2 + m_2] \ }
(j_1, j_2, m_1+1, m_2-1 \vert j m )_{q} \cr
}
$$

\aab

In the limiting case where $q = 1$, the
recursion relations 4 to 12 are in agreement
with the results of Kibler and Grenet [12].

\aa

\noindent {\it 3.2. Recursion relations for $U_q(su_{1,1})$}

\aa

14. The action of ${\cal J}_{\pm}$
on $\vert k_1 \mp \de,
          k_2 \pm \de, j, \kappa \ra_q $
leads to

$$
\eqalign
{
& \sqrt{ [j \mp k_1 \pm k_2 + 1]   \
         [j \pm k_1 \mp k_2    ] } \
(k_1 k_2 \kappa_1 \kappa_2 \vert j \kappa )_{q} \cr
 = & \ q^{{1 \over 2} (\pm \kappa_1 \mp \kappa_2 - k_1 + k_2) }
 \sqrt{ [\kappa_1 \pm k_1] \ [\kappa_2 \pm k_2 \pm 1] } \
(k_1 - \de, k_2 + \de, \kappa_1 \mp \de, \kappa_2 \pm \de \vert j \kappa )_{q}
\cr
 + & \ q^{{1 \over 2} (\mp \kappa_1 \pm \kappa_2 - k_1 + k_2) }
 \sqrt{ [\kappa_2 \mp k_2] \ [\kappa_1 \mp k_1 \mp 1] } \
(k_1 + \de, k_2 - \de, \kappa_1 \mp \de, \kappa_2 \pm \de \vert j \kappa )_{q}
\cr
}
$$

15. The action of ${\cal K}_{\pm}$ on
    $\vert k_1, k_2, j, \kappa \mp 1 \ra_q $
    leads to

$$
\eqalign
{
& \sqrt{ [\kappa \mp j \mp 1] \ [\kappa \pm j] } \
(k_1 k_2 \kappa_1 \kappa_2 \vert j \kappa )_{q} \cr
=
 \pm & q^{{1 \over 2} (\pm \kappa_1 \pm \kappa_2 - k_1 + k_2) }
 \sqrt{ [\kappa_1 \pm k_1] \ [\kappa_2 \mp k_2 \mp 1] } \
(k_1 - \de, k_2 + \de, \kappa_1 \mp \de, \kappa_2 \mp \de
                       \vert j, \kappa \mp 1 )_{q}
\cr
 \mp & q^{{1 \over 2} (\mp \kappa_1 \mp \kappa_2 - k_1 + k_2) }
 \sqrt{ [\kappa_2 \pm k_2] \ [\kappa_1 \mp k_1 \mp 1] } \
(k_1 + \de, k_2 - \de, \kappa_1 \mp \de, \kappa_2 \mp \de
                       \vert j, \kappa \mp 1 )_{q}
\cr
}
$$

16. The action of ${\Lambda}_{\pm}$ on
    $\vert k_1 \mp \de,
           k_2 \mp \de, j, \kappa \ra_q $
    leads to

$$
\eqalign
{
& \sqrt{ [j \pm k_1 \pm k_2 \pm 1  ]   \
         [j \mp k_1 \mp k_2 + \ump ] } \
( k_1 k_2 \kappa_1 \kappa_2 \vert j \kappa )_{q} \cr
& = q^{+{1 \over 2} (k_1 + k_2 - \kappa_1 + \kappa_2 + \ump) }
 \sqrt{ [\kappa_1 - k_1 - \hmp] \ [\kappa_1 + k_1 + \hmp] } \
( k_1 \mp 1, k_2, \kappa_1, \kappa_2 \vert j \kappa )_{q} \cr
& + q^{-{1 \over 2} (k_1 + k_2 + \kappa_1 - \kappa_2 + \ump) }
 \sqrt{ [\kappa_2 - k_2 - \hmp] \ [\kappa_2 + k_2 + \hmp] } \
( k_1, k_2 \mp 1, \kappa_1, \kappa_2 \vert j \kappa )_{q} \cr
}
$$

\aa

The recursion relations 14 to 16 are in accordance with
the results by Nomura [14],
   Groza {\it et al.} [15],
  Kachurik and Klimyk [15], and
               Aizawa [16]
who derived recursion relations for the Clebsch-Gordan
coefficients of $U_q(su_{1,1})$
from $q$-deformed hypergeometric functions.

\bb

\centerline {\bf Acknowledgements}

\aa

One of the authors (Yu.F.~S.) is grateful to the {\it Institut de Physique
Nucliaire de Lyon} for the hospitality extended to him during his stay
in Lyon-Villeurbanne where this work was completed.

\vfill\eject

\centerline {\bf References}

\aa

\noindent
\item{[1]}
Biedenharn L.C.,
J.~Phys.~A: Math. Gen.  {\bf 22} (1989) L873.

\aab

\noindent
\item{[2]}
Macfarlane  A.J.,
J.~Phys.~A: Math. Gen.  {\bf 22} (1989) 4581.

\aab

\noindent
\item{[3]}
Chakrabarti  R. and Jagannathan  R.,
J.~Phys.~A: Math. Gen.  {\bf 24} (1991) L711.

\aab

\noindent
\item{[4]}
Schirrmacher  A., Wess  J. and Zumino  B.,
Z.~Phys.~C {\bf 49} (1991) 317.

\aab

\noindent
\item{[5]}
Smirnov  Yu.F. and Wehrhahn  R.F.,
J.~Phys.~A: Math. Gen.  {\bf 25} (1992) 5563.

\aab

\noindent
\item{[6]}
Kibler M.R., in {\it Symmetry and Structural Properties of Condensed
Matter}, Eds.,\break
Florek W., Lipinski D. and Lulek T.,
World Scientific, Singapore (1993), p. 445.

\aab

\noindent
\item{[7]}
Dore\v si{\'c} M., Meljanac S. and Milekovi{\'c} M.,
Fizika B {\bf 2} (1993) 43.

\aab

\noindent
\item{[8]}
Barbier R., Meyer J. and Kibler M.,
Preprint LYCEN 9348.

\aab

\noindent
\item{[9]}
Schwinger J., Report U.S. Atom. Ener. Comm. NYO-3071 (1952), published in
{\it Quantum Theory of Angular Momentum}, Eds., Biedenharn L.C.
and van Dam H., Academic Press, New York (1965).

\aab

\noindent
\item{[10]}
Smorodinski{\u \i} Ya.A. and Shelepin L.A.,
Usp. Fiz. Nauk {\bf 106} (1972) 3.
(English translation: Usp. Fiz. Nauk. {\bf 15} (1972) 1.)

\aab

\noindent
\item{[11]}
Rasmussen W., J. Phys. A: Math. Gen. {\bf 8} (1975) 1038.

\aab

\noindent
\item{[12]}
Kibler M. and Grenet G., J. Math. Phys. {\bf 21} (1980) 422.

\aab

\noindent
\item{[13]}
Biedenharn L.C. and Louck J.D., in
{\it Encyclopedia of Mathematics and Its Applications},
Ed., Rota G.-C., Addison-Wesley, Reading, Massachusetts
(1981), Vol. 8/9.

\aab

\noindent
\item{[14]}
Nomura M., J. Phys. Soc. Japan {\bf 59} (1990) 1954~;
Nomura M., J. Phys. Soc. Japan {\bf 59} (1990) 2345.

\aab

\noindent
\item{[15]}
Groza V.A., Kachurik I.I. and Klimyk A.U., J. Math. Phys. {\bf 31} (1990)
2769~;
\break
Kachurik I.I. and Klimyk A.U., J. Phys. A: Math. Gen. {\bf 24} (1991) 4009.

\aab

\noindent
\item{[16]}
Aizawa N., J. Math. Phys. {\bf 34} (1993) 1937.

\aab

\noindent
\item{[17]}
Smirnov Yu.F. and Kibler M.R., in
{\it Symmetries in Science VI: From the Rotation Group to Quantum Algebras},
Ed., Gruber B., Plenum Press, New York (1993), p. 691.

\aab

\noindent
\item{[18]}
Moshinsky M. and Quesne C., Phys. Lett. {\bf B 29} (1969) 482~;
Moshinsky M. and Quesne C., J. Math. Phys. {\bf 11} (1970) 1631.

\bye